\documentclass[a4paper,12pt]{article}

\usepackage[english]{babel}
\usepackage[a4paper,tmargin=3truecm,bmargin=3truecm,rmargin=2.5truecm,
lmargin=2.5truecm,twoside,verbose=true]{geometry}

\usepackage{cancel,graphicx}
\usepackage[dvips]{hyperref}

\usepackage{amsmath,amssymb}

\numberwithin{equation}{section}

\newcommand{\be}{\begin{equation}} 
\newcommand{\ee}{\end{equation}} 
\newcommand{\bea}{\begin{eqnarray}} 
\newcommand{\eea}{\end{eqnarray}} 

\label{Schr}

\newcommand {\rd} {{\rm d}}
\newcommand {\p} {\partial}

\allowdisplaybreaks[1]

\renewenvironment{thebibliography}[1]
         {\section*{References}\frenchspacing\small
          \begin{list}{[\arabic{enumi}]}
         {\usecounter{enumi}\parsep=2pt\topsep 0pt
         \settowidth{\labelwidth}{[#1]}
         \leftmargin=\labelwidth\advance\leftmargin\labelsep
         \rightmargin=0pt\itemsep=1pt\sloppy}}{\end{list}}

\title{Construction of potentials using
 mixed scattering data\footnote{e-mail: lassaut@ipno.in2p3.fr.}}
\author{M. Lassaut$^{a,b}$, S.Y. Larsen$^c$, 
S.A. Sofianos$^b$, J.C. Wallet$^d$}
\date{}

\begin{document}

\maketitle
\vspace*{-1cm}
\begin{center}
\textit{ $^{a}$ Institut de Physique Nucl\'eaire}\par
\textit{CNRS/Universit\'e Paris-Sud 11 (UMR8608),F-91406 Orsay Cedex, 
        France\\}\par
\vskip 0,2 true cm
\textit{$^b$ Physics Department, University of South Africa, 
       Pretoria 0003, South Africa}\par
\vskip 0,2 true cm
\textit{$^c$ Department of Physics, Temple University, Philadelphia, PA 19122, USA}\par
\vskip 0,2 true cm 
\textit{$^d$ Laboratoire de Physique Th\'eorique, B\^at.\ 210}\par
\textit{CNRS/Universit\'e Paris-Sud 11 (UMR8627),  
       F-91405 Orsay Cedex, France\\}\par
\vskip 0,2 true cm
\textit{}\par

\end{center}

\vskip 2cm

\begin{abstract}
 The long-standing problem of constructing a potential from 
mixed scattering data is discussed.  We first consider the fixed-$\ell$ 
inverse scattering problem. We show that the zeros of the regular 
solution of the Schr\"odinger equation, $r_{n}(E)$ which are 
monotonic functions of  the energy, determine a 
unique potential when the domain of energy is such that the 
$r_{n}(E)$'s range from zero to infinity. The latter method is 
applied to the domain 
  $\{E \geq E_0, \ell=\ell_0 \} \cup \{E=E_0, \ell \geq \ell_0 \}$ 
 for which  the zeros of the regular solution are monotonic in both 
parts of the domain and still range from zero to infinity.
 Our analysis suggests that  a unique potential can  be obtained 
from the mixed scattering data 
   $\{ \delta(\ell_0,k), k \geq k_0 \} \cup
\{ \delta(\ell,k_0), \ell \geq \ell_0 \}$
 provided that certain  integrability conditions required for the fixed $\ell$-problem, are fulfilled. The uniqueness is demonstrated 
using the JWKB approximation. 
\end{abstract}

\pagebreak
\section{Introduction}
Three-dimensional (3-D) inverse quantum scattering problem amounts to reconstruct some interaction potential from a set of experimentally accessible scattering data. For spherically symmetric potentials{\footnote{Notations: $\vec{r}=(r\sin\theta\cos\phi, r\sin\theta\sin\phi, r\cos\theta)$, $r>0$, $0\le\theta\le\pi$, $0\le\phi\le2\pi$.}} $V(r)$,
 the case that will be considered throughout  this paper, it is known \cite{newt} that the 3-D problem 
$\left(- \Delta_{\vec{r}}+V(r) \right) \psi(E,\vec{r}) =E  \psi(E,\vec{r})$ ($\Delta_{\vec{r}}$ is the 3-D Laplacian, $E$ is the energy) 
can be reduced, 
thanks to the partial wave decomposition 
$\psi(E,\vec{r})= \sum_{l\in\mathbb{N}}\sum_{m=-\ell}^{m=\ell}\frac{\psi_{\ell}(E,r)}{r} Y_{\ell,m}(\theta,\phi)$ (the $Y_{\ell,m}$'s are the spherical harmonic functions) to the study of the radial Schr\"odinger equation with centrifugal barrier term depending on the angular momentum $\ell$. This latter{\footnote{The potential $v(r)$ is normalised such that ${{2m}\over{\hbar^2}}V(r)=v(r)$ where $m$ is the mass for the particle. We use throughout this paper the units \cite{CS} $\hbar=2m=1$; then, $V(r)=v(r)$ and the energy $E$ verifies $E=k^2$ where $k$ is the wave number.}} is given by 
\begin{align}
\left(\frac{\rd^2}{\rd r^2}\right. +\left. 
   E-V(r) -\frac{(\ell + 1/2)^{2}-1/4}{r^2} \right)
     \psi_{\ell}(E,r) =  0 
\label{eq:Schr}
\end{align}
where $ \psi_{\ell}(E,r)$ is called the regular solution which is uniquely defined, as usual \cite{newt,CS}, by 
the Cauchy condition $\lim_{r \to 0} \psi_{\ell}(E,r) r^{-\ell -1}=1$. It behaves for positive values of $E$ as 
$\psi_{\ell} \propto \sin(k r - \ell \pi/2 + \delta(\ell,k))$ when  $r\to\infty$ ($k=\sqrt{E}$, see second footnote), provided that $V(r)$ satisfies the integrability condition \cite{CS}
\begin{align}
 \int_b^{+\infty} \vert V(r) \vert \rd r  <  \infty, \quad b>0, \qquad
            \int_0^{\infty} r  \vert V(r) \vert \rd r  <  \infty
\label{eq:int}
\end{align}
Here, the $ \delta(\ell,k)$'s are the phase shifts, the relevant scattering data for the present paper. Notice that the whole scheme can be extended to real values for $\ell$ verifying $2\ell+1 > 0$ \cite{newt}. \par 

Basically, approaches to 3-D inverse scattering can be 
classified in two categories \cite{newt,CS},
namely, those using scattering data for fixed values of 
either the energy $E$ or the angular momentum $\ell$. In the first category, the so-called fixed-$E$ inverse scattering problem, 
Loeffel  \cite{Loef68}  obtained theorems predicting a unique
potential from the knowledge of the phase-shifts  $\delta(\ell,k)$ at
a specific energy  $E=k^2$, for all (non-discrete) non negative values 
of $\lambda=\ell+1/2$. When the set of data is reduced to discrete 
values of $\lambda=\ell+1/2$ for non-negative integer $\ell$, Carlson's 
theorem \cite{Carlson}  predicts a unique potential   $V(r)$, provided that this latter belongs to a suitable class \cite{CS,Loef68}. 
The  Newton series permits one to construct the potential $V(r)$ from the knowledge of the discrete partial wave 
scattering data \cite{Loef68}. In the second category,   the so-called fixed-$\ell$ inverse scattering problem,  the potential $V(r)$
 that satisfies \eqref{eq:int} can be constructed 
from the phase-shifts $\delta(\ell,k)$ given for 
all momenta $k\in (0,\infty)$  and from the discrete spectrum data 
(bound state energies and the corresponding normalization constants). For details, see \cite{CS}.\par 

Apart from the above two distinct  approaches there exists also a possibility of constructing a potential
starting from input scattering data that are partly $E$-dependent and partly
$\ell$-dependent.  The idea of exploiting simultaneously $E$- and 
$\ell$-dependent scattering data was first explored by Grosse and 
Martin \cite{GM}  for the construction of confining potentials.
They conjectured that the knowledge of the ground state energies 
$E_{\ell}^{(0)}$, for all non-negative integers $\ell$, 
allows one to recover the potential in a unique way.
The problem of reconstructing potentials from bound states has been studied numerically in Ref. \cite{YIL}. In \cite{RZ}, Rudyak and Zakhariev have constructed potentials from $E$- and $\ell$-dependent data,  however assuming that $a E + b \ell (\ell+1)$ is a constant. In this case, the corresponding analysis is based on an extension of the Newton's method.

In this paper, we construct a potential obtained from input scattering data that are partly $E$- and 
$\ell$-dependent, when no extension of the Newton's method is available. Let us summarize the results of our study. First we show that, in the fixed-$\ell$ problem, the knowledge of the zeros of the regular solution the Schr\"odinger equation permits one to determine the potential. More precisely, we show that  the zeros of the regular solution, denoted hereafter by $r_{n}(E)$, $n\geq 1$, which are known to be monotonic functions of the energy, whatever the behavior of the potential may be \cite{Sturm}, determine a unique potential provided the domain of the energy is such that $r_{n}(E)$ range from $0$ to $\infty$. Note that the reconstruction of the potential from the zeros of the regular solution enters the category of inverse nodal problems introduced by Hald and McLaughlin \cite {HL}. As a second result, we show that piecewise constant potentials can be constructed from only one single line of zeros still ranging from $0$ to $\infty$, say $r_{n_0}(E)$, for some fixed $n_0$. As a third result of this paper, we prove uniqueness theorems, stating basically that two different potentials satisfying \eqref{eq:int} cannot have a common line of zeros ranging from $0$ to $\infty$. In the second part of the analysis, we turn to the mixed $\ell$ and $E$ problem and apply the above results to the domain $\{E \geq E_0, \ell=\ell_0 \} \cup \{E=E_0, \ell \geq \ell_0 \}$ for which the zeros of the regular solution for the Schr\"odinger equation are monotonic functions on both parts of this domain and still range from zero to infinity. This domain cannot be dealt with (any extension of) the Newton's method. The first part of the analysis suggests that a unique $\ell$- and $E$-independent potential could be obtained from the set of mixed scattering data given by $\{ \delta(\ell_0,k), k \geq k_0 \} \cup\{ \delta(\ell,k_0), \ell \geq \ell_0 \}$. As the last result of this paper, we show that this is indeed true within the semi-classical JWKB approximation, provided there is only one turning point. Note that the uniqueness theorem 1 still applies for the mixed data, leading to the construction of piecewise constant potentials from the zeros of the regular solution of the Schr\"odinger equation. General details on the JWKB approximation and its full applicability can be found e.g in \cite{engin} while specific applications to the Schr\"odinger equation may be found e.g in  \cite{newt, CS}  . Basically, it provides a way to determine a semi-classical expansion (in $\hbar$) of the solution of the Schr\"odinger equation and is valid whenever the potential changes slowly during an oscillation of this solution.\par 

Our analysis has natural applications in heavy ion physics as well as general properties of scattering processes although applications in other areas, such as acoustics and/or geophysics seem likely to be possible. In heavy ion (elastic) scattering physics, a natural application of the present paper concerns the problem of discrete ambiguities encountered in optical model analysis \cite{goldberg} and would at least provide a new insight on past works of this problem \cite{Sabatier, Cuer}. We have undertaken the corresponding study which will be reported elsewhere \cite{LW}. The present analysis can be easily extended 
to more complicated set of (mixed) scattering data. In particular, we emphasize that the most noticeable feature of our approach is that it can be applied to inverse scattering problems not reducible to the $\ell=0$-fixed inverse scattering problem through a Liouville transformation (defined for instance in \cite{CS}), as it is the case for the fixed-$E$ problem for potentials of finite range \cite{Loef68} or for the case treated by Rudyak and Zakhariev \cite{RZ} also for potentials of finite range. Note that, when the inverse scattering problem is reducible to the usual $s$-wave inverse scattering problem, there exists an extension of the Newton's method available  for the scattering problem considered.\par 

The paper is organised as follows. The section \ref{sec2} involves the whole analysis. In subsection \ref{s21} , we consider the fixed-$\ell$ problem. In subsection \ref{ssmix}, we discuss in detail the mixed scattering problem. The subsection \ref{sjwkb} deals with the JWKB approximation applied to the mixed scattering data. In section \ref{sec3}, we discuss the results and conclude.\par 

\section{Formalism}\label{sec2}
\subsection{The fixed-$\ell$ inverse scheme\label{s21}}
We assume that the potential fulfills the integrability condition \eqref{eq:int}. Let us consider the regular solution of the Schr\"odinger equation $\psi_{\ell}(E,r)$ for fixed positive energy $E$ and fixed $\ell$, a function of $r\ge 0$. It can be realized that $\psi_{\ell}(E,r)$  and $\psi'_{\ell}(E,r)$ cannot vanish at the same value(s) of $r$ for $2 \ell+1 >0$ (except for $r=0$) and $\ell >0$ as shown in \cite{Hille} so that $\psi_{\ell}(E,r)$ can only have simple zeros (except at $r=0$). Owing to the asymptotic properties of $\psi_{\ell}(E,r)$ and ${{\psi_{\ell}}\over{\psi'_{\ell}}}(E,r)$ as $r \to \infty$, it can be further realized that the set of zeros is countable. According to these observations, these zeros (the origin $r=0$ being excluded) can then be naturally ordered starting from the smallest value that will be denoted by $r_1(\ell,E)$. A generic zero of the regular solution will be denoted by $r_n(\ell,E)$, $n\in\mathbb{N}$, $n\ge1$. This defines the numbering of the zeros that is used in the following analysis. Notice that, thanks to the fact that all the zeros are simple, one has the following (strict) ordering $0< r_1(\ell,E) < r_2(\ell,E) < \cdots < r_n(\ell,E)< \cdots $. \par 
For any potential, the zeros satisfy the monotonicity   properties,
i) $ E=k^2 \mapsto r_n(\ell,E)$ is a decreasing function,
as  has been shown by Sturm in 1830's \cite{Sturm} and ii)
$\ell \mapsto r_n(\ell,E)$ is an increasing function. This can be easily shown. More precisely,  we have
\bea
\label{zero1}
       \frac{\partial}{\partial E} r_n(\ell,E) 
       &= & -\int_0^{r_n(\ell,E)} {\rm d}r' \psi_{\ell}(E,r')^2 /
            \left(\frac{\partial}{\partial r} 
           \psi_{\ell}(E,r_n(\ell,E) \right)^2 
\\
           \frac{\partial}{\partial \ell} r_n(\ell,E) 
          &= & (2 \ell+1) \ \int_0^{r_n(\ell,E)} {\rm d}r' 
             \frac{\psi_{\ell}(E,r')^2}{r'^2} \  /
             \left(\frac{\partial}{\partial r} 
             \psi_{\ell}(E,r_n(\ell,E) \right)^2
\label{zero2}
\eea
Note that, as mentioned above, $\psi_{\ell}(E,r)$ and $(\partial \psi_{\ell}/\partial r)(E,r)$ cannot vanish simultaneously 
for $2 \ell+1 >0$, except for $r=0$ and $\ell >0$ \cite{Hille}, so that the denominators in (\ref{zero1})
and (\ref{zero2}) cannot vanish. \par 
For potentials satisfying  (\eqref{eq:int}) 
the function $E \mapsto r_n(\ell,E)$  is such that $r_n(\ell,E)\to 0$ 
for $E\to\infty$ \cite{Chadan67}. This is a consequence of
equation (I.5.6) of \cite{CS}. In the absence of bound states  
$ r_n(\ell,E)\to\infty$ for $E\to0$. \par 
For negative or zero values of $E$, the regular solution of the Schr\"odinger equation has zeros provided that the potential has bound states,
 and that $E  > E_1$, where $E_1 < 0$
denotes the ground state energy\footnote{We adopt this notation to denote 
the ground state by $E_1$ instead of the traditionally used $E_0$
in order to be consistent with the meaning given to $n$, namely 
to denote zeros of the wave function}.  
The number of zeros is finite for values of the energy $E$ such that  $E_1 < E \leq 0$. For details see \cite{Chadan67}. When the potential has $N$ bound states,  $E_1 < E_2 ... < E_N$, 
if $n \leq N $, 
we have $ r_n(\ell,E)\to\infty$ for  $E\to E_n$
\cite{Chadan67}, 
whereas for $n > N$, $ r_n(\ell,E)\to\infty$ as  $E\to0$.
This is illustrated in the Fig. {\ref{f1} where we  have drawn the 
first  four zeros of the regular solution
for a Bargmann transparent potential of Ref. \cite{LLSR}, which has 
a bound state at the energy $E=-1$ in $1/L^2$ units. All zeros, but the first one, go to infinity as $E$ goes to $0$.
The  first zero, however, goes to infinity   when $E \to E_1=-1$.
\begin{figure}
\centering
\includegraphics[width=8.cm, height=6cm]{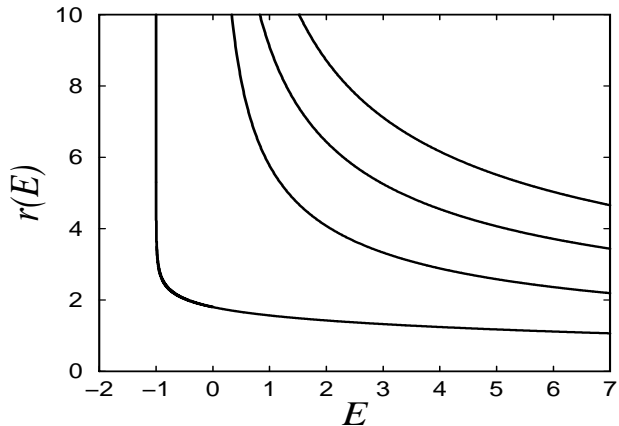}
\caption{\label{f1}
Zeros $r_n(E),\ n=1,2,3$, and $4$, of the regular 
s-wave solution for a Bargmann potential  (for more details
see text).}
\end{figure}
In the fixed-$\ell$ inversion, if all  the zeros are assumed to 
be known, i.e. $E \mapsto r_n( \ell,E)$ is known and $r_n(\ell,E)$ 
describes the entire interval $[0,\infty[$, then
the potential $V_\ell(r)$ is uniquely determined.
This can be easily checked. For this we 
consider the  Sturm-Liouville 
problem on $[0,R]$, i.e., the equation
\be
       \psi_{\ell}''(r)+\left(E-V_{\ell}(R-r) 
       -\frac{\ell(\ell+1)}{(R-r)^2} \right)  \psi_{\ell}(r)=0,
\label{stur}
\ee
coupled with the Dirichlet conditions
$$
       \psi_{\ell}(0)=\psi_{\ell}(R)=0\,.
$$ 
The spectral data are the eigenvalues $E_n^*$'s, 
such that $R=r_n(\ell,E_n^*)$,
and  the normalization  constants \cite{Sacks}
$$
       \rho_n=\frac{\int_0^R \psi_{\ell}(r')^2 \rd r'}
      {\psi'_{\ell}(0)^2} \, . 
$$
The latter are given by the positive values 
$$
      \rho_n=-\frac{\rd}{\rd E} r_n(\ell,E) \vert_{E=E_n^*} \, .
$$
The potential, assumed to be  square integrable on $[0,R]$ is uniquely 
determined on $[0,R]$ by the spectral data $\{ E_n^*,\rho_n\} $ \cite{GL}.  
The technique of constructing $V$ from this set of data  is 
well-known \cite{CS,Sacks,CM,CKM}.

If we consider now the Sturm Liouville problem on $[0,R]$ consisting of
the equation   
\be
    \psi_{\ell}''(r)+\left(E-V_{\ell}(r)-\frac{\ell(\ell+1)}{r^2} 
     \right) \psi_{\ell}(r)=0,
\label{stur1}
\ee
coupled with the Dirichlet conditions
$$\psi_{\ell}(0)=\psi_{\ell}(R)=0$$ 
we recover the Gel'fand-Levitan approach 
at the limit $R\to\infty$ \cite{CS,HO} .

For the purpose of using mixed, $E$-  and $\ell$-dependent data,
we start with the following remark. For a fixed $\ell$, the nth zero of the
regular solution can be considered as a function of energy $r_n(\ell,E)$,
which is monotonic
 and such that $\psi_{\ell}(E,r_n(\ell,E)) \equiv 0$.
It defines a line of zeros as we mentioned in the introduction. Moreover  $E \mapsto r_n(\ell,E)$ 
 admits an inverse function $r \mapsto E_{n,\ell}(r)$ which is also
 monotonic and is the  inverse of this line of zeros.
For example, consider the potential $V \equiv 0$  in the S-wave ($\ell=0$).
The regular solution is proportional to 
$\sin(\sqrt{E} r )$. The lines of zeros are given by 
$r_n(0,E)=n \pi/\sqrt{E}$ and the inverses of these lines are 
$E_{n,0}=n^2 \pi^2/r^2$.

Suppose that the inverses $r \mapsto  E_{n,\ell}$,  of 
all lines of zeros $n \geq 1$, are known for $r$
 running from $0$ to $\infty$. This implies  that 
 $E_n^*=E_{n,\ell}(R)$ is also known and that the $\rho_n$ are 
 given by 
$$
  \frac{1}{\rho_n}=-\frac{\rd }{\rd r} E_{n,\ell}(r) \vert_{r=R}
$$
We can then construct the desired potential, assumed to be 
locally square integrable on $[0,R]$. If only  the  inverse  $r \mapsto  E_{n,\ell}$ of a single 
line  of zeros is known, no method is available to recover the potential, 
except in the special case of piecewise constant potentials.
What we can show is a uniqueness property, if 
$E_{n_1,\ell}(r)$ is known for all positive $r$ (see
the theorems given below).\par 
In the special case of piecewise constant potentials, having 
discontinuities at values of $r = a_j$, $j =1,\ldots ,j_{max}$, 
and being zero for $r > a_{j_{max}}$, it is easy to show that there 
is a one to one correspondence  between the discontinuities of the third
derivative of $E_{n_1,\ell}(r)$ with respect to $r$ 
and  the discontinuities of $V$.
This suggests the following lemmas:\\
\noindent{\bf Lemma 1:}
{\it For a piecewise constant potential, the knowledge of a single line of
 zeros $r(E)=r_n(\ell,E)$ ($n$ fixed) allows
the reconstruction of the potential in a unique way provided that the latter  function 
has values ranging from zero to infinity when $E \in ]0,+\infty[$. }\\

The following variant of the lemma 1, although useless for the mixed set of data considered in the next subsection,   works also for potentials having bound states. 
Denoting  $r \mapsto E(r)$  the  inverse   function of the function $E \mapsto r(E)$ which is  the single 
line of zeros $r(E)$ considered, we state that \\

\noindent{\bf Lemma 2:}
{\it For a piecewise constant potential, the knowledge of  the inverse of a single line of
 zeros $E(r)=E_{n,\ell}(r)$  allows the reconstruction of
the potential in a unique way provided that the latter  function of the variable $r$ is known on the whole interval $]0,+\infty[$}.\\
The proof uses the arguments given above,  in particular the monotonicity property of the line with respect to the energy, and the third derivative of
$$          
  \psi(E(r),r) \equiv 0 .
$$
To be more specific, if $E'''$ has a discontinuity at $r=a$  then
\be
        \frac{\rd^3 E}{\rd r^3}(a^+)-\frac{\rd^3 E}{\rd r^3}(a^-)
         =-2 \frac{\rd E}{\rd r}(a)
           \left [V(a^+)-V(a^-)\right] \ .
\label{disc1}
\ee
This is equivalent to the relation
\be
      \frac{\rd^3 r}{\rd E^3}(E_a^+)-\frac{\rd^3 r}{\rd E^3}(E_a^-)  
           =2\left(\frac{\rd r}{\rd E}(E_a)\right)^3 
          \left[V(r(E_a^+))-V(r(E_a^-))\right]
\label{disc2}
\ee
Since the potential is zero for $r> a_{j_{max}}$, the 
relation (\ref{disc1}) allows us to reconstruct the potential 
between $a_{j_{max}}$ and $a_{j_{max}-1}$. The
procedure can be repeated at each $a_j$, and the potential is obtained at
$r \neq a_j$  by summing the successive values at each
discontinuity appearing beyond $r$.

As an illustration consider  the  potential 
\be 
 V(r)  =\begin{cases}
   -2  & r < 2 \cr
    -1 & 2 < r < 3 \cr
     0 & r> 3\cr
\end{cases}
\label{pcpot}
\ee
In  Fig. {\ref{f2}, we have drawn the function 
$r \mapsto -E'''(r)/(2 E'(r))$, 
related to the inverse  $E(r)$ of the first 
line of zeros for the $S$-wave regular solution  of the Schr\"odinger 
equation with  the potential (\ref{pcpot}).
\begin{figure}[th]
\centering
\includegraphics[width=8cm, height=6cm]{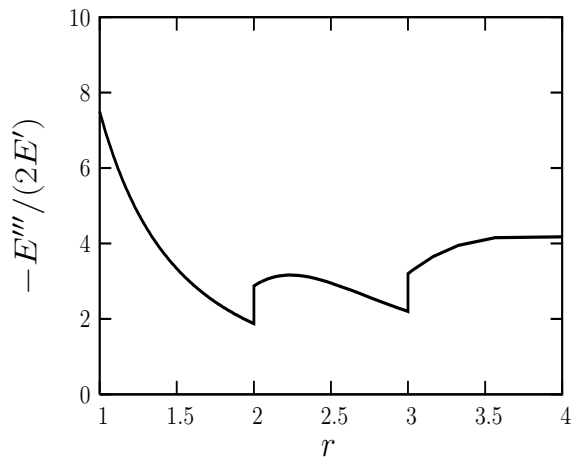}
\caption{\label{f2}
The structure of the function $-E'''(r)/(2 E'(r))$ 
related to the inverse  $E(r)$ of the first 
line of zeros for the $S$-wave regular solution 
corresponding to the potential (\ref{pcpot}).}
\end{figure}
Clearly, the discontinuities of  $r \mapsto -E'''(r)/(2 E'(r))$ happen at 
the points where $V$ has discontinuities,  namely $r=2$ and $r=3$,
and the equation (\ref{disc1}) is satisfied. We know that the
potential is zero beyond $r=3$. 
 So we have $0=V(3^+)$.
 From the curve the discontinuity  $V(3^+)-V(3^-)$ is equal to $1$ then $V(3^-)=-1$.
As $V$ is piecewise constant, it is constant on $]2,3[$ and    we have $V(2^+)=V(3^-)=-1$.
From the curve the discontinuity 
$V(2^+)-V(2^-)$ is equal to $1$  so that $V(2^-)=-2$.
We thus  recover the potential  (\ref{pcpot}).\par 
This method cannot be applied to a potential defined by a continuous function.
Nevertheless, for such potentials, the following uniqueness 
theorems hold:\\
\noindent{\bf Theorem 1:} 
{\it Consider  two  potentials $V_1$ and $V_2$ satisfying \eqref{eq:int} and locally constant in the vicinity of zero. 
 For the fixed-$\ell$ problem, assume that both potentials have no bound states. Suppose that two integers 
$n_1$ and $n_2$ exist such that the $n_1$-th line of zeros for 
the regular solution for $V_1$  coincides with the $n_2$-th line of  
zeros for the regular solution for $V_2$,
 ($r_{n_1}(\ell,E) \equiv r_{n_2}(\ell,E)$) with both 
lines having values ranging from $0$ to $\infty$ as the energy $E$ varies in $]0,\infty[$. 
Then $V_1\equiv V_2$. }\\

Its variant, still useless for our mixed set of data,  but working for potentials having bound state yields: \\

\noindent{\bf Theorem 2:} 
{\it Consider  two  potentials $V_1$ and $V_2$ satisfying \eqref{eq:int}
and locally constant in the vicinity of zero. 
 For the fixed-$\ell$ problem, assume that two integers 
$n_1$ and $n_2$ exist such that the inverse of the $n_1$-th line of zeros for 
the regular solution for $V_1$  coincides with the inverse of the $n_2$-th line of  
zeros for the regular solution for $V_2$ on the whole  interval $]0,+\infty[$,
 $ ((\forall r > 0) E_{n_1,\ell}(r)  \equiv E_{n_2,\ell}(r) ).  $
Then $V_1\equiv V_2$. }\\
Let us prove the theorem 1. The proof of its  variant, the theorem 2, quite similar to that of theorem 1, is not reported here. \\
First let $\psi_{1}(E,r)$ and $\psi_{2}(E,r)$
(we suppress the index $\ell$ for convenience)  be the regular solutions 
of the Schr\"odinger equation for potentials $V_{1}$ and $V_{2}$, 
respectively, constrained by the Cauchy conditions 
$\lim_{r \to 0} \psi_{i}(E,r) r^{-\ell-1}=1,\  i=1,2$ in the vicinity
 of zero. For an $S$-wave, this is equivalent to $\psi_i'(E,0)=1,\
 i=1,2$, the prime being the derivative with respect to $r$.
Setting $ \Delta V=V_1-V_2$ and noting that the Wronskian is zero 
for $r=0$ we obtain
\bea
     W(\psi_1,\psi_2)&=&\psi_1'(E,r) \psi_2(E,r)- \psi_1(E,r) \psi_2'(E,r) 
\nonumber\\
 &= &\int_0^r {\rm d}r' 
  \ \Delta V(r') \psi_1(E,r') \psi_2(E,r') \  .
\label{Wronskian}
\eea
For $r=r(E)$ (we adopt the convention  $r_{n_1}(\ell,E)=r_{n_2}(\ell,E)=r(E)$)
we have
\be
    (\forall E)  \qquad\quad \int_0^{r(E)} {\rm d}r'  \ \Delta V(r')
          \psi_1(E,r') \psi_2(E,r') =0 \ . 
\label{eqnf}
\ee
The latter integral exists since $\psi_i,\ i=1,2$ are continuous and 
both potential $V_1$ and $V_2$  satisfy \eqref{eq:int}) and are locally 
constant in the vicinity of zero.

 Differentiating twice Eq. (\ref{eqnf}) with respect to $E$,  
which is feasible as the first and the second derivative of 
$   \psi_1(E,r) \psi_2(E,r)$ with respect to the energy $E$  
is a function continuous of the variable  $r$ \cite{RN},  
and taking 
into account that  $\psi_{i}(E,r(E))=0, \ i=1,2$ for every value of 
$E$, we obtain  

\be
  (\forall E)\qquad   \int_0^{r(E)} {\rm d}r'   \  \Delta V(r')
 \  \frac{\p^2}{\p E^2} \left[ \psi_1(E,r') \ \psi_2(E,r') \right]  =0 
%
\label{eqnf1}
\ee
We recall that, due to its monotonicity, $E \mapsto r(E)$ admits an
 inverse $r \mapsto E(r)$. We define the Kernel
%
\begin{equation}
         K(r,r')=\frac{\p^2}{\p E^2} 
        \left[ \psi_1(E(r),r') \ \psi_2(E(r),r') \right]\,,\qquad
             r' \leq r \ .
\label{Kernel}
\end{equation}
The diagonal part,  $K(r,r)$,  equal to   
\be
       K(r,r)=2 \frac{\p }{\p E}  \psi_1(E(r),r) \,
       \frac{\p }{\p E}  \psi_2(E(r),r)= 2 
      \left( \frac{ \rd r}{dE} \right)^2   \frac{\p }{\p r}  
       \psi_1(E(r),r) \, \frac{\p }{\p r}  \psi_2(E(r),r) \, .  
\label{diagker}
\ee
is never zero, since $\psi_i$ and $\frac{\rd}{\rd r} 
\psi_i$ ($i=1,2$) do not vanish simultaneously, 
except for $r=0$ and for $\ell >0$ \cite{Hille}.

Introducing the definition (\ref{Kernel}), 
the equation (\ref{eqnf1}) is equivalent to  
\be
(\forall r \geq 0)  \qquad\quad \int_0^r dr'  K(r,r') \  \Delta V (r') =0  \ .
\label{eqnf2}
\ee 

Since the potential $V_1$ ($V_2$) is  locally constant 
in the vicinity of zero, there exist $\tilde V_1$ and 
 $\epsilon_1$ ( $\tilde V_2$ and $\epsilon_2$) 
such that 
\bea
(\forall r \leq \epsilon_1)  \qquad\quad V_1(r) & = & \tilde V_1  \nonumber\\
(\forall r \leq \epsilon_2) \qquad\quad V_2(r) & = & \tilde V_2
\label{cons} 
\eea
In the vicinity of zero, the $n_1$-th line of zeros of the regular 
solution for $V_1$ is given by 
\be
               r_{n_1}(\ell,E)=\frac{j_{\ell,n_1}}{\sqrt{E-\tilde V_1}}
\ee
where $j_{\ell,n_1}$ is the $n_1$-th zero, except for the origin, of the
function $x \mapsto j_\ell(x)$ involving the 
spherical Bessel function
$j_\ell(x)$ \cite{erd}.\\

This implies  that
\be
     \frac{\rd \ln r_{n_1}(\ell,E)}{\rd E}
                =- \frac{1}{2} \ \frac{1}{E-\tilde V_1}\,.
\label{zerd}
 \ee

Taking into account (\ref{cons}) and (\ref{zerd}) as well as the 
identity $r_{n_1} \equiv r_{n_2}$  we see that for 
$r \leq \epsilon=\inf(\epsilon_1,\epsilon_2) $ 
we have $\tilde V_1-\tilde V_2=0$. Therefore   $\Delta  V(r)$ 
is zero for $r \leq \epsilon$.

Differentiating (\ref{eqnf2}) with respect to $r$, ( possible since the derivative $r' \mapsto K_r(r,r')$ is continuous) and taking into account 
that $\Delta V$ is zero for $r \leq \epsilon$, leads to the 
Volterra equation
\be
   (\forall r \geq \epsilon) \qquad\quad   \Delta V(r)+\int_{\epsilon}^r 
      \rd r'  K_1(r,r') \Delta V(r') =0 
\label{eqnf3}
\ee 
where 
\be
      K_1(r,r')=\frac{ K_r(r,r')}{K(r,r)}, \qquad\quad K_r(r,r')=
         \frac{\partial K}{\partial r}(r,r')
\label{K1}
\ee
Let us consider (\ref{eqnf3}) for $r \in [\epsilon,R]$. The Kernel $K_1$ is
bounded and  continuous.  This is due to the fact that  $K(r,r)$ 
is continuous and never vanishes on 
the  compact interval  $[\epsilon,R]$. 
 So $\vert K(r,r) \vert $ is bounded from below 
by a strictly positive value. Thus Eq. (\ref{eqnf3}) has a unique 
 solution for $r \in [\epsilon,R]$ \cite{RN}, which reads 
$\Delta V(r) =0$. By increasing $R$ we can conclude that 
$\Delta V(r) =0$ for every $r$.
This terminates the proof. Note that the fact that $K_1(r,r)$ 
behaves like $1/r$ in the vicinity of 
zero forbids us to conclude that (\ref{eqnf3}) has a unique
solution for $\epsilon=0$ \cite{cham}.This is the reason why 
the potentials have been assumed to be locally constant 
in the vicinity of zero. 
\subsection{Inversion scheme from mixed data}
\label{ssmix}

Consider the set  
$\{ E \geq E_0, \ell=\ell_0 \}  \cup \{ E=E_0,\ell \geq \ell_0 \}$.
Similarly to Sect. \ref{s21}, we  may define lines of zeros
in which the n-th line of zeros $r_n(\ell,E)$ of the regular 
solution describes a line formed of two parts. 
 In the first part,  the zeros $r_n(\ell_0,E)$ 
 range from $r=0$, ($E=\infty$),
 to $r_0=r(\ell_0,E_0), (E=E_0)$, as the energy 
$E$ varies from $\infty$ to $E_0 \ (\ell_0 $ being fixed); 
in the second part, $r_n(\ell,E_0)$  
 has values ranging from $r_0=r(\ell_0,E_0), 
(\ell=\ell_0)$, to $\infty$, $(\ell=\infty)$ as $\ell$ ranges
from $\ell_0$ to $\infty$ ($E_0$ being fixed). 
This has been  verified 
for potentials  satisfying the integrability 
conditions \eqref{eq:int}  \cite{CS}.
The monotonicity property required in the above  lemma 1 and  theorem 1
is preserved in both domains. \\

The lemma $1$ is still valid  for piecewise constant potentials. In the special case where a discontinuity  $v=V(r_0^-)-V(r_0^+)$
appears at the junction point between both parts of the domain, i.e. at $r_0=r_n(\ell_0,E_0)$, the value of $v$ can be determined as follows. One first adds $v$ to the bare value $V_0$ of the potential at the origin, obtained  by adding all discontinuities (except $v$). Then, for $r$ close to the origin, the zeros are given by
\be
               r_n(\ell_0,E)=\frac{j_{\ell_0,n}}{\sqrt{E-V_0-v}}
\label{rnz}
\ee
 and the value of $v$ can be obtained from the relation 
\be
     \frac{\rd \ln r_n(\ell_0,E)}{\rd E}
          =- \frac{1}{2} \ \frac{1}{E-V_0-v}\, . 
\ee 
To demonstrate the applicability of the theorem 1,
we consider potentials satisfying the integrability 
conditions  \eqref{eq:int} and locally constant in the vicinity of zero.
Using the Wronskian relation 
(\ref{Wronskian}),  we have to show that,  if 
\be
      \int_0^{r(\ell_0,E)} \rd r'  \, \Delta V(r')
      \psi_1(\ell_0,E,r') 
      \psi_2(\ell_0,E,r') = 0 \qquad { \rm for} \ \    E \geq E_0  \ ,
\label{eqnfm}
\ee
 and 
\be
      \int_0^{r(\ell,E_0)} \rd r'  \, \Delta V(r') \psi_1(\ell_,E_0,r') 
      \psi_2(\ell,E_0,r') = 0 \qquad {\rm for} \ \ \ell \geq \ell_0 \ , 
\label{eqnfmm}
\ee
then $V_1 \equiv V_2$. 
(Here we have put as previously  $r_{n_1}=r_{n_2}=r$.)

This can be proved in two steps. First, for $r \leq r_0=r(\ell_0,E_0)$, 
Eq. (\ref{eqnfm}) is zero for every $E \geq E_0$.
Consequently,  $V_1 \equiv V_2$ for $r \leq r_0$ when $V_1,V_2$  
are locally constant in the vicinity of zero.

Secondly,  taking into account the above 
argument, Eq. (\ref{eqnfmm})  can be rewritten  
\be
    \int_{r(\ell_0,E_0)}^{r(\ell,E_0)} \rd r'  \Delta V(r')
             \psi_1(\ell_,E_0,r') \psi_2(\ell,E_0,r')=0  
\label{eqnfmmp}
\ee
Differentiating twice with respect to $\ell$, which is feasible as  the first and the second 
derivative of  $ \psi_1(\ell_,E_0,r) \psi_2(\ell,E_0,r)$ with respect to $\ell$ is continuous with respect to the variable $r$ \cite{RN}, 
 and inverting the monotonic 
function $ \ell  \mapsto r(\ell,E_0)$,  denoted by $r \mapsto \ell(r,E_0)$, 
we obtain
\bea
     (\forall r \geq r_0) \  & & \int_{r_0}^{r} \rd r'  \, \Delta V(r')
            K(r,r') =0  
\label{eqnfmmp1} \\
     (\forall r' \leq r)  & & K(r,r')=
     \frac{\partial^2}{\partial \ell^2} 
      [\psi_1(\ell(r,E_0),E_0,r') \psi_2(\ell(r,E_0),E_0,r')] \ . 
\nonumber
\eea 

The diagonal part  $K(r,r)$  is equal to 
\bea
     K(r,r) &=&2 \frac{\p }{\p \ell}  \psi_1(\ell(r,E_0),E_0,r) \,
     \frac{\p }{\p \ell}  \psi_2(\ell(r,E_0),E_0,r) \nonumber\\
     &=& 2 \left( \frac{ \rd r}{d\ell} \right)^2   \frac{\p }{\p r}  
           \psi_1(\ell(r,E_0),E_0,r) \, 
            \frac{\p }{\p r}  \psi_2(\ell(r,E_0),E_0,r) \ ,
\label{diagkerlp}
\eea
and does not vanish for $r \geq r_0 >0$. 

Differentiating Eq. (\ref{eqnfmmp1}) with respect to $r$, (possible as $ r' \mapsto K_r(r,r')$ is continuous)  we obtain the 
Volterra equation
\be
   (\forall r \geq r_0) \qquad\quad   \Delta V(r)+\int_{r_0}^r 
      \rd r'  K_1(r,r') \Delta V(r') =0 
\label{eqnf3l}
\ee 
where $K_1$ is given by (\ref{K1}).
Consider now Eq.(\ref{eqnf3l}) for $r \in [r_0,R]$. 
Since $K(r,r)$  in Eq. (\ref{diagkerlp}) is continuous on $[r_0,R]$, 
its absolute value reaches its  strictly positive  minimum $m$. 
The Kernel $K_1(r,r')$ is then bounded and continuous and   
the equation (\ref{eqnf3l}) has a unique solution $\Delta V(r)=0$ on $[r_0,R]$.
Increasing $R$, we conclude that 
$ \Delta V(r)=0$ for every $r \geq r_0$  and then  $V_1 \equiv V_2$ on 
the whole half axis.

We have shown that a single line of zeros,  which, 
for the data considered, always has values ranging from zero to infinity and moreover is
monotonic, determines the potential uniquely. 
The remaining question is to
examine whether  the set of  mixed data
\be
      \{ \delta (\ell=\ell_0,k) \ \ k \in [k_0, + \infty[ \} 
          \cup \{ \delta(\ell,k_0) \ell \in [\ell_0,+\infty[ \}
\label{specd}
\ee
 associated to the set $\{\ell=\ell_0, k \in [k_0,\infty[ \} 
\cup \{k=k_0, \ell \in [\ell_0,\infty[\}$,
 determines a line of zeros, and thus the potential, in a unique way -
 which is suggested by the analogy with the $\ell$-fixed problem. 
In the absence of bound states, all lines of zeros 
$E \mapsto r_n(\ell,E)$, monotonic with respect to $E$, 
range from zero ($E$ infinite) to infinity ($E=0$) when $E$ has values ranging from 
infinity to zero. In this case, we know that the potential,  
when it satisfies \eqref{eq:int}, is recovered in a unique way, 
given the phase-shifts $\delta(\ell,k)$ for all $k \geq 0$.
The condition  \eqref{eq:int} 
excludes all pathologies; for example, potentials behaving  asymptotically
like $1/r^2$, encountered in particular in the presence 
of a zero energy bound state, or ghost components in the Jost 
function \cite{LLSR}. With the mixed data we are in the same 
situation, namely all the lines of zeros 
are monotonic and range from zero ($E$ infinite, $\ell=\ell_0$)
 to infinity ($E=E_0$, $\ell$ infinite). 
Thus we expect that (\ref{specd}) is associated with a 
unique potential decreasing faster than $1/r^2$ at infinity.
We cannot prove this in the general case, 
but we can investigate the problem in a JWKB approach.

Note that a way to treat  the general case is to set 
a self-consistent procedure. First we note that,  in order to calculate the phase-shift say $\delta(\ell,k)$, there is no need to know the potential for distances  
smaller than $r_n(\ell,E)$, whatever is $n$. It is sufficient to consider the Schr\"odinger equation on the interval $[r_n(\ell,E), \infty[$  
because only the ratio $\psi_{\ell}/\psi'_{\ell}$, where the prime denotes the derivative with respect to $r$, has to be known to determine the phase-shift, 
In this respect,  we first determine the potential 
in terms of the phase-shift $\delta(\ell,k_0) (\delta(\ell_0,k))$ 
and the  n-th ($n$ fixed for all the procedure) zero of the 
regular solution by solving the Schr\"odinger equation 
on $[r_n(\ell,k_0),\infty[ \   ([r_n(\ell_0,k),\infty[)$
because the phase-shift does not depend on the potential 
$V(r)$ for $r \leq r_n(\ell,k_0) (r_n(\ell_0,k))$.
The algebraic construction of the potential is guaranteed 
by  the fact that the monotonicity property of the  n-th line of zeros of 
the regular solution  in terms of the energy does not depend  on the potential.
 Then we have to determine the zeros.
For instance take the zeros of the free solution $V \equiv 0$ 
then determine $V$ then recalculate the zeros etc.. 

We first consider results based on the Born approximation.
In the seventies Reignier \cite{Reignier} used a Born approximation 
of the scattering amplitude to show that the knowledge 
of the phase-shift at a fixed energy, $E_0=k_0^2$ say, for each 
integer $\ell$ is equivalent to the knowledge of the 
Fourier sine transform of the potential $r V(r)$,
\be
     g(q)= \int_0^{\infty} \sin(q r) r V(r) \rd r\,,
\ee
 for $q \leq 2 k_0$. 
The scattering amplitude is determined from the phase-shifts 
at fixed energy $\delta(\ell,k_0)$ for $\ell=0,1,2,\ldots$, 
$E=E_0=k_0^2$.

Generally, the integral is assumed to be 
zero for $q >2 k_0$  \cite{CS,Sabatier} leading 
to potentials 
\be
         r V(r)= \frac{2}{\pi} \ \int_0^{2 k_0} \sin(q r) \,g(q)\, \rd q
\ , \label{b1} 
\ee
such that $r V(r)$ is an entire function of $r$ of order 1. 
Other extensions of $g(q)$  are studied in  \cite{Sabatier04}. 

More recently Habashy and Wolf  \cite{HW} have studied the 
reconstruction of a potential, having compact support and 
spherical symmetry, from its 3D-Fourier transform throughout the 
Ewald limiting sphere,  $\vert k_0 (\vec{u}-\vec{u'}) \vert$ where 
$\vec{u},\vec{u'}$
are normalized to unity and take all possible directions.
This is equivalent to the Fourier transform of the potential for 
values  of $\vec{k}$
such that  $\  \vert \vec{k} \vert  \leq 2 k_0$.
In this case, the inverse sine transform $r \tilde V(r)$ 
 coming from Eq. (\ref{b1})  is given in terms of the potential $V(r)$ by 
\be
        r \widetilde V(r) =\int_{-R}^R \rd r' \ V(r') 
           \frac{\sin[2 k_0 (r-r')]}{\pi (r-r')}\,, 
             \qquad -R \leq r \leq R\,,
\ee
where $R$ denotes the support of $V$.
For negative values of $r$, the authors of \cite{HW}  have 
put $V(-r)=V(r)$. The latter equation can be solved numerically,
thanks to the spectrum of the kernel involved in its right hand side, 
namely,  to the $\nu_n \in ]0,1[$, \ $\psi_n \in L^2([-1,1]) $ 
(depending on $k_0$) such that 
\be
       \int_{-1}^1 \rd x'  \frac{\sin[2 k_0 (x-x')]}{\pi (x-x')}
       \  \psi_n(x')=\nu_n \ \psi_n(x)\,,
       \qquad  -1 \leq x \leq 1 \ .
\ee
Coming back to Eq. (\ref{b1}), a possible way to extend our knowledge of
$g(q)$ beyond $2 k_0$  is to take the Born approximation for the missing
phase shifts $\delta(\ell=0,k)$ for $k \geq  k_0$. This is given by
\be
       \int_0^{\infty} \sin(k r)^2 \ V(r) \rd r =- k \delta(\ell=0,k)
\ . \ee
The derivative with respect to $k$ yields
\be
 g(q)=  \int_0^{\infty} \sin(q r)  r V(r) \rd r =- \frac{\rd 
      (k \delta(\ell=0,k))}{\rd k}\,,\qquad
           \forall q=2 k \geq 2 k_0\,.
\ee
This implies that $q(q)$, known for $q \leq 2 k_0$, is now known 
for every positive $q$, and that $V(r)$ is uniquely given by 
\be
         r V(r)= \frac{2}{\pi} \int_0^{\infty} \sin(q r) \,g(q)\, \rd q
\, . 
\label{b2}
\ee
Consequently, the knowledge of 
 $\{\delta(\ell,k_0), \ell \in {\mathcal N} \} \cup 
 \{\delta(\ell=0,k),k \geq  k_0 \}$ 
allows us to recover the potential in Born approximation 
if $k_0$ is sufficiently high. 
\subsection{The JWKB approximation}\label{sjwkb}
In what follows, we examine to what extent the mixed data 
(\ref{specd}) allow us  to recover the  potential in the semi-classical JWKB approximation. General details on the JWKB approximation and its full applicability can be found e.g in \cite{engin} while specific applications to the Schr\"odinger equation may be found e.g in  \cite{newt, CS}. It has also been extensively used in the fixed energy inverse scattering 
(see, for example,  \cite{Kuj,Sabatier,Cuer,Fied} and 
references  therein). Basically, we recall that this semi-classical approximation provides a way to determine a semi-classical expansion (in $\hbar$) of the solution of the Schr\"odinger equation and is valid whenever the potential changes slowly during an oscillation of this solution.\par 
In the JWKB approximation, the phase-shift is given by
\be
       \delta(\ell,k)=\lim_{r \to \infty} 
        \bigg(\int_{r(\lambda,k)}^r K(\lambda,k,r')  \rd r'
 - \int_{r_{{\rm free}}(\lambda,k)}^r  
            K_{{\rm free}}(\lambda,k,r') \bigg)
       \rd r'
\label{phase}
\ee
(here $E=k^2$ in $1/L^2$ units and $\lambda = \ell + 1/2$). 
In the absence of a Coulomb component, we have
\be
      K(\lambda,k,r)=\sqrt{k^2-V(r)-\lambda^2/r^2}\,, \qquad\quad      
      K_{{\rm free}}(\lambda,k,r)=\sqrt{k^2-\lambda^2/r^2}
\, . 
\ee
In (\ref{phase}),  $r(\lambda,k)$ is the turning point 
relative to the function $K$ considered, assumed here to 
be unique for the sake of simplicity.  More precisely  
$r(\lambda,k)$ is a solution of the equation   $K(\lambda,k,r)=0$.

For the part of the spectrum 
$\delta(\lambda,k_0), \ \lambda \geq \lambda_0$ 
we use the results of  Sabatier \cite{Sabatier} and Cuer 
\cite{Cuer} for a single turning point.
 We assume that ($k=k_0$ being fixed) for every $\lambda$,
 the equation 
\be
          k_0^2-V(r)-\lambda^2/r^2=0
\label{turning}
\ee
has a unique solution denoted by $r(\lambda,k_0)$. This happens when 
the function $r \mapsto g(r)=r^2 (k^2-V(r))$ is monotonic.
For $V \equiv 0$ the equation (\ref{turning}) has a unique solution, 
denoted by $r_{{\rm free}}(\lambda,k)=\lambda/k$.
In \cite{Sabatier,Cuer}  the phase-shift is given by:
\be
       \delta(\ell,k_0)=\int_{\lambda}^{\infty} 
       \rd \lambda' \ \sqrt{\lambda'^2-\lambda^2} \ 
       \frac{\rd}{\rd \lambda'} \ln \left( 
      \frac{r(\lambda',k_0}{r_{{\rm free}}(\lambda',k_0)} 
      \right)
\label{delta}
\ee
where $V(r)$ is assumed to be differentiable  to ensure the 
derivative of the turning point with respect to $\lambda$ exists.
The function 
\be
          \lambda \mapsto g(\lambda)=\lambda  \frac{\rd}{\rd \lambda} 
           \ln \left( \frac{r(\lambda,k_0}{r_{{\rm free}}
           (\lambda,k_0)} \right)
\ee
is assumed to be Lebesgue integrable.

From Eq. (\ref{delta}) we have 
\be
        \frac{\rd}{\rd \lambda} \delta(\ell,k_0)=-\lambda \ 
        \int_{\lambda}^{\infty} \rd \lambda' \ 
         \frac{1}{\sqrt{\lambda'^2-\lambda^2}} \ 
        \frac{\rd}{\rd \lambda'} \ln \left( 
        \frac{r(\lambda',k_0}{r_{{\rm free}}(\lambda',k_0)} 
        \right)
\label{deltap}
\ee
the derivation being possible for $\lambda \ne 0$  as 
the integrand is dominated on $]\lambda,\infty[$ by 
the integrable function  
$\lambda' \mapsto \vert g(\lambda') 
\vert /(\lambda' \sqrt{\lambda'^2-\lambda^2}).$

Using the Lebesgue-Fubini theorem \cite{godement,rudin}  
which works when 
$\rd \delta(\ell,k_0)/\rd \lambda \in L^1({\cal R})$ and bounded, 
as made for  the Abel transform, the turning point is given in terms 
of the phase-shifts by
\be
        \ln \left( \frac{r(\lambda,k_0)}{r_{{\rm free}}
        (\lambda,k_0)}\right) = \frac{2}{ \pi} \int_{\lambda}^{+\infty} 
     \frac{\rd \delta(\ell'=\lambda'-1/2,k_0)}{\rd \lambda'}
        \frac{1}{\sqrt{\lambda'^2-\lambda^2}}   \rd \lambda'
\label{rec}
\ee
provided that the logarithm is zero for $\lambda$ infinite. The above hypothesis that $\rd \delta(\ell,k_0)/\rd \lambda $ is  bounded avoids pathologies in the integration  in the vicinity of $\lambda$. 
For a single turning point we recover the result 
of Loeffel: the knowledge of $\delta(\ell,k)$ for 
every $\lambda=\ell+1/2$ positive allows us to recover the potential.
Indeed,  for $\lambda$ infinite the logarithm tends to zero and 
$r(\lambda,k_0)$ tends to infinity. 
When $\lambda$ tends to zero  the equation (\ref{turning}) 
has as unique solution $r=0$ when  $\lim_{r \to 0} V(r) r^2=0$.

For $r \geq r_0=r(\lambda_0,k_0)$, the potential is given  by
\be
V(r(\lambda,k_0))=k_0^2-\frac{\lambda^2}{r(\lambda,k_0)^2}=
k_0^2 \left[1-\left( \frac{r_{{\rm  free}}(\lambda,k_0)}{r(\lambda,k_0)} 
\right)^2 \right] \ . \ee
Note that the behavior of the potential for $r(\lambda,k_0)$ or 
$\lambda$ infinite is related to the behavior of the 
phase-shift $\delta(\ell,k_0)$ for $\lambda$ infinite.

For $r \leq r_0$, we introduce the turning point $r=r(\lambda_0,k)$ 
as the  solution of 
\be
                k^2-V(r)-\frac{\lambda_0^2}{r^2}=0
\label{turningp}
\, . 
\ee
We assume that the latter equation has a unique solution. 
This happens when $r \mapsto h(r)=V(r)+\lambda_0^2/r^2 $
 is monotonic, for example for $V \geq 0$ or for
 $V$ attractive but ``weak'' enough when compared to the centrifugal
 barrier, in the sense that 
$ \vert \vert r^2 V'(r) \vert \vert_{\infty} <  \lambda_0^2$ .
The aforementioned  function $h$ is positive
and infinite at the origin and zero for $r = \infty$. Thus,  
we are sure to have at least one solution. 
The monotonicity of $h$ implies the  uniqueness of the 
solution of the equation $h(r)=k^2$, denoted by $r(\lambda_0,k)$. It implies also the monotonicity of 
$k \mapsto r(\lambda_0,k)$, which is 
 monotonically decreasing  with respect to $k$. 
For $k$ infinite $r(\lambda_0,k) \to 0$.

For $V \equiv 0$  Eq. (\ref{turningp}) has a unique solution, 
denoted by $r_{{\rm free}}(\lambda_0,k)=\lambda_0/k$. The equation 
(\ref{phase}) is rewritten as
\be
            \delta(\ell_0,k)= \int_{0}^{k} 
              \sqrt{k^2-k'^2}  \frac{\rd }{\rd  k'} 
         \bigg( r(\lambda_0,k')
         -r_{{\rm  free}}(\lambda_0,k') \bigg) \rd k' \ .
\ee
When $V$ is differentiable, the derivative of the turning point 
with respect to $k$ exists. We assume that the latter  is 
locally integrable.
The above equation is derived with respect to $k$ and thus
\be
      \delta'_k(\ell_0,k)=\frac{\rd }{\rd k}  
           \delta(\ell_0,k)=  \int_{0}^{k} 
            \frac{k }{\sqrt{k^2-k'^2}} \frac{\rd }{\rd  k'} 
           \bigg( r(\lambda_0,k')-r_{{\rm free}}(\lambda_0,k') 
           \bigg) \rd k'
\, . 
\ee
Using the procedure of the inverse Abel transform \cite{RN,Abel} 
we obtain
\be
\int_0^{k} \frac{ \ \delta'_k(\ell_0,u)}{\sqrt{k^2-u^2}}
          \rd u=\int_0^k \frac{\rd }{\rd  k'} 
           \bigg( r(\lambda_0,k')-r_{{\rm  free}}(\lambda_0,k') 
           \bigg) \rd k' \int_{k}^{k'} \rd u  
   \frac{u}{\sqrt{(u^2-k^2)(k'^2-u^2)}}
\ee
using the Lebesgue-Fubini theorem \cite{godement},\cite{rudin} which works when $\delta'_k(\ell_0,k)$ is Lebesgue integrable and bounded. 
Once the integration over $u$ is performed we obtain
\be
  r(\lambda_0,k)-r_{{\rm free}}(\lambda_0,k) 
             = \frac{2}{  \pi }
          \int_0^{k} \frac{\ \delta'_k(\ell_0,k')}{\sqrt{k^2-k'^2}}
          \rd k'
\label{invk} \, , 
\ee
provided that $ r(\lambda_0,k)-r_{{\rm free}}(\lambda_0,k) $ is zero 
for $k=0$, which happens for finite range potential.
In this latter case, for $k$ high enough, 
$ r(\lambda_0,k) =\lambda_0/k$ and tends to zero for $k$ infinite.

The potential 
$$
       V(r(\lambda_0,k))=k^2-\frac{\lambda_0^2}{r(\lambda_0,k)^2}
$$
can be determined from (\ref{invk}) for $0 \leq r \leq R=r(\lambda_0,k_0)$,
i.e. for $k \geq k_0$. However,  Eq. (\ref{invk}) requires knowledge of the phase-shift $\delta(\lambda_0,k)$ for every value 
of $k$, whereas they are known only for $k \geq k_0$.

Nevertheless, these phase-shifts for 
$k \leq k_0$  can be determined  from the potential 
beyond the distance $R$,   $R=r(\lambda_0,k_0)$.
Indeed,    for $k \leq k_0$, (\ref{phase}) can be written as
\be
     \delta(\ell_0,k)=\int_{\lambda(k)}^{+\infty}
        \sqrt{k^2-V(r(\lambda,k_0))-
          \frac{\lambda_0^2}{r(\lambda,k_0)^2} }
          \frac{\rd r(\lambda,k_0)}{\rd \lambda}  \ \rd \lambda 
  -   \int_{\lambda_0  k_0/k}^{+\infty} 
          \sqrt{k^2-\frac{k_0^2 \lambda_0^2}{\lambda^2} }
            \frac{1}{k_0}    \rd \lambda \,.
 \ee
Here, $\lambda(k)$ corresponds to the value of $\lambda$ 
for which the first square root  in the previous equation 
vanishes, i.e it is a solution of
\be
        k^2-k_0^2 + \frac{\lambda^2-\lambda_0^2}{r(\lambda,k_0)^2} =0\,. 
\ee
This implies, for $k \leq k_0$, that
\be
      \delta(\ell_0,k) = \int_{\lambda(k)}^{+\infty} 
      \sqrt{k^2-k_0^2 +
      \frac{\lambda^2-\lambda_0^2}{r(\lambda,k_0)^2}}
      \frac{\rd r(\lambda,k_0)}{\rd \lambda} \  \rd \lambda
      -\int_{k_0 \lambda_0/k}^{+\infty}  \sqrt{k^2 
      -\frac{ k_0^2 \ \lambda_0^2}{\lambda^2}}
      \frac{1}{k_0} \rd \lambda\,,
\label{deltm} 
\ee
with $r(\lambda,k_0) $ given by (\ref{rec}) in terms of the phase-shifts
$\delta(\lambda,k_0), \lambda \geq \lambda_0$.
\section{Conclusion}\label{sec3}
In the present work, we have been concerned with a non-standard inverse scattering problem, namely with the construction of 
the potential from scattering data information which involves the use of phase-shifts 
$$
      \{ \delta(\ell_0,k), k \geq k_0 \} \cup
       \{ \delta(\ell,k_0), \ell \geq \ell_0 \}
$$
 corresponding to the domain 
$$ \{(\ell_0,k),  k \geq k_0 \} \cup
       \{ (\ell,k_0), \ell \geq \ell_0 \}
$$
and without requiring any extension  of Newton's method for  inverse scattering problems. \par 
First, we have shown that, in the fixed-$\ell$ problem, the knowledge of the zeros of the regular solution the Schr\"odinger equation permits one to determine the potential: in fact, the zeros of the regular solution $r_{n}(E)$, $n\geq 1$, determine a unique potential provided the domain of the energy is such that $r_{n}(E)$ has values ranging from $0$ to $\infty$. As a second result, we have shown that piecewise constant potentials can be constructed from only one single line of zeros (say $r_{n_0}(E)$, for some fixed $n_0$) with values
still ranging from $0$ to $\infty$. We also proved uniqueness theorems, stating basically that two different potentials satisfying \eqref{eq:int} cannot have a common line of zeros ranging from $0$ to $\infty$. Furthermore, we have also considered the mixed $\ell$ and $E$ problem and applied the above results to the domain $\{E \geq E_0, \ell=\ell_0 \} \cup \{E=E_0, \ell \geq \ell_0 \}$ for which the zeros of the regular solution for the Schr\"odinger equation are monotonic functions on both parts of this domain and have values still ranging from zero to infinity. This domain cannot be dealt with any extension of the Newton's method. A unique $\ell$- and $E$-independent potential could be obtained from the set of mixed scattering data given by $\{ \delta(\ell_0,k), k \geq k_0 \} \cup\{ \delta(\ell,k_0), \ell \geq \ell_0 \}$. As the last result, we have shown that this is indeed true within the semi-classical JWKB approximation, provided there is only one turning point. \\
Furthermore, we have shown that in Born approximation 
the following mixed scattering data 
$$
      \{\delta (\ell=0,k),   k \in [k_0, + \infty[ \}
            \cup \{ \delta(\ell,k_0), \ell \in 
            {\cal N} \}
$$
lead to an unique potential, still assumed to be $\ell$- and 
$E$-independent, which is the inverse Fourier sine transform of a 
function deduced from the data.

To conclude, our method, which does not use an extension of 
Newton's method, takes advantage of a set of mixed scattering data.

A natural application  concerns the discrete ambiguities  (i.e. the set of different
families of potentials with practically the same fixed energy phase-shift) encountered in heavy ions elastic scattering
optical model analyses \cite{goldberg}. These latter ambiguities
have been studied within the JWKB approximation in Refs. \cite{Sabatier,Cuer}.

Finally we point out that the flexibility of our method,  based upon the properties of the zeros of the regular solution, provide a way to study inverse scattering problems which are not  reducible to the $\ell=0$-fixed inverse scattering problem though a Liouville transformation. This latter class of problems involves the inverse scattering problem from our mixed set of data,  for which there is no expected available extension of the Newton's method.\par

\section*{Acknowledgments} 

We are grateful to R.J. Lombard  for many discussions and
a careful reading of the manuscript.
One of us (ML) is very grateful to the University of South Africa
 for its kind hospitality.


\end{document}